\documentclass[12pt]{article}
  \usepackage{amsfonts}
  \usepackage{amsmath}
\usepackage{amssymb}
\usepackage{amscd}
\usepackage[dvips]{graphicx}
\begin{document}

~~
\bigskip
\bigskip
\begin{center}
{\Large {\bf{{{Twist deformation of  doubly
enlarged Newton-Hooke Hopf algebra}}}}}
\end{center}
\bigskip
\bigskip
\bigskip
\begin{center}
{{\large ${\rm {Marcin\;Daszkiewicz}}$}}
\end{center}
\bigskip
\begin{center}
\bigskip

{ ${\rm{Institute\; of\; Theoretical\; Physics}}$}

{ ${\rm{ University\; of\; Wroclaw\; pl.\; Maxa\; Borna\; 9,\;
50-206\; Wroclaw,\; Poland}}$}

{ ${\rm{ e-mail:\; marcin@ift.uni.wroc.pl}}$}

\end{center}
\bigskip
\bigskip
\bigskip
\bigskip
\bigskip
\bigskip
\bigskip
\bigskip
\bigskip
\begin{abstract}
We provide fifteen twist-deformed doubly enlarged Newton-Hooke quantum space-times. In $\tau$ approaching infinity limit the twisted doubly enlarged
Galilei spaces are obtained as well.
\end{abstract}
\bigskip
\bigskip
\bigskip
\bigskip
\eject

\section{{{Introduction}}}

Recently, in article \cite{Gomis:2008jc} there has been proposed the so-called classical doubly
enlarged Newton-Hooke Hopf algebra $\;{\mathcal U}_{0}(\widehat{\widehat{NH}}_{\pm})$. It contains, apart from rotation $(M_{ij})$,
boost $(K_{i})$, acceleration-like $(F_i)$ and space-time translation $(P_{i}, H)$ generators,
the additional ones denoted by $R_{i}$.  Its algebraic part looks as follows\footnote{Present in
the above commutation relations   parameter $\tau$ denotes
the characteristic for Newton-Hooke algebra cosmological time scale.}
\begin{eqnarray}
&&\left[\, M_{ij},M_{kl}\,\right] =i\left( \delta
_{il}\,M_{jk}-\delta _{jl}\,M_{ik}+\delta _{jk}M_{il}-\delta
_{ik}M_{jl}\right)\;\; \;, \;\;\; \left[\, H,P_i\,\right] =\pm
\frac{i}{\tau^2}K_i
 \;,  \notag \\
&~~&  \cr &&\left[\, M_{ij},K_{k}\,\right] =i\left( \delta
_{jk}\,K_i-\delta _{ik}\,K_j\right)\;\; \;, \;\;\;\left[
\,M_{ij},P_{k }\,\right] =i\left( \delta _{j k }\,P_{i }-\delta _{ik
}\,P_{j }\right) \;,\nonumber
\\
&~~&  \cr &&\left[ \,M_{ij},H\,\right] =\left[ \,K_i,K_j\,\right] =
\left[ \,K_i,P_{j }\,\right] =0\;\;\;,\;\;\;\left[ \,K_i,H\,\right]
=-iP_i\;\;\;,\;\;\;\left[ \,P_{i },P_{j }\,\right] = 0\;,\label{nnnga}\\
&~~&  \cr &&\left[\, F_i,F_j\,\right] =\left[\, F_i,P_j\,\right]
=\left[\, F_i,K_j\,\right] =0\;\; \;, \;\;\;\left[\,
M_{ij},F_{k}\,\right] =i\left( \delta _{jk}\,F_i-\delta
_{ik}\,F_j\right) \;,\nonumber\\
&~~&  \cr
&&\left[\, R_i,R_j\,\right] =\left[\, R_i,P_j\,\right]
=\left[\, R_i,K_j\,\right] =\left[\, R_i,F_j\,\right] =0\;\; \;, \;\;\;\left[\,
M_{ij},R_{k}\,\right] =i\left( \delta _{jk}\,R_i-\delta
_{ik}\,R_j\right) \;,\nonumber\\
&~~&  \cr &&~~~~~~~~~~~~~~~~~~~~~~~\left[\,
H,F_{i}\,\right] =2iK_i  \;\;\;,\;\;\;      \left[\,
H,R_{i}\,\right] =3iF_i\;,\nonumber
\end{eqnarray}
while the coproduct and antipode take the form
\begin{eqnarray}
\Delta_0(a) = a\otimes 1 +1\otimes a\;\;\;,\;\;\;S_{0}(a) =-a\;.\label{cop}
\end{eqnarray}
with $a$ = $M_{ij}$, $K_{i}$,  $F_i$, $P_{i}, H$ and $R_{i}$. Besides,
the corresponding symmetry transformations of  nonrelativistic space-time
  are given by\footnote{Here we take the following assignements of the parameters:\\
\begin{tabular}{lll}
           $a_i$ &- & spatial translations (generators $P_i$)
\cr
           $v_i$ &- & boosts (generators $K_i$)
\cr
           $b_i$ &- &  accelerations  (generators $F_i$)
\cr
           $\alpha_{ij}$ &- &$O(d)$ space rotations (generators
           $M_{ij}$)
\cr
           $a_0$ & - &time translation (generator $H$)
\cr
           $c_i$ & - &new (additional) parameters (generators $R_i$).
           \end{tabular}}
\begin{eqnarray}
&x_i& \longrightarrow \;\;\alpha_{ij} x_j+ a_i C_{\pm} \left(\frac{t}{\tau}\right) + v_i  \tau S_{\pm} \left( \frac{t}{\tau}\right) \pm 2 b_i  \tau^2
\left(C_{\pm} \left(\frac{t}{\tau}\right)  - 1\right) + \nonumber\\
&~~&~~~~~~\,\pm\;6c_i\tau^3\left(S_{\pm} \left(\frac{t}{\tau}\right)  - \frac{t}{\tau}\right)\;,\label{trans1}\\
&t& \longrightarrow \;\;t+a_0\;. \label{trans2}
\end{eqnarray}
with $C_{+} [\frac{t}{\tau}] = \cosh \left[\frac{t}{\tau}\right]$,
$C_{-} [\frac{t}{\tau}] = \cos \left[\frac{t}{\tau}\right]$, $S_{+}
[\frac{t}{\tau}] = \sinh \left[\frac{t}{\tau}\right]$, $S_{-}
[\frac{t}{\tau}] = \sin \left[\frac{t}{\tau}\right]$, for finite cosmological constant $\tau$, and
\begin{eqnarray}
&x_i& \longrightarrow \;\;\alpha_{ij}x_j+ a_i + v_i  t +  b_i  t^2+ c_it^3\;,\label{trans3}\\
&t& \longrightarrow \;\;t+a_0\;, \label{trans4}
\end{eqnarray}
in the case of parameter $\tau$ approaching infinity.

It should be noted, that the Hopf structure (\ref{nnnga}), (\ref{cop}) is the largest known explicitly symmetry (quantum)  group
at nonrelativistic level. By its different contraction schemes we get respectively: \\

{ i)} For $R_i \to 0$ - the acceleration-enlarged  Newton-Hooke Hopf algebra $\;{\mathcal U}_{0}({\widehat{NH}}_{\pm})$ proposed in \cite{Gomis:2008jc},
\cite{lucky0}, \\

{ ii)} In the case of $R_i$ and $F_i$ generators approaching zero - the (usual) Newton-Hooke quantum group $\;{\mathcal U}_{0}({{NH}}_{\pm})$
 \cite{bacry}, \\

{ iii)} For $R_i \to 0$ and $\tau \to \infty$ - the acceleration-enlarged  Galilei Hopf structure $\;{\mathcal U}_{0}({\widehat{G}})$ provided in
\cite{lucky1}, \\

{iv)} For both $R_i$ and $F_i$ operators approaching zero as well as for parameter $\tau$ running to infinity - the (usual) Galilei quantum group  $\;{\mathcal U}_{0}({{G}})$, \\
\\
and, finally \\

{ v)} In the case of $\tau \to \infty$ - the (new) doubly enlarged  Galilei Hopf algebra $\;{\mathcal U}_{0}(\widehat{{\widehat{G}}})$\;. \\
\\

In this article we consider the Abelian twist-deformations (see \cite{twist1}-\cite{twist3}) of doubly enlarged  Newton-Hooke Hopf structure $\;{\mathcal U}_{0}(\widehat{{\widehat{NH}}}_{\pm})$. In such a way, in accordance with the contraction schemes { i)}-{iv)}, we rediscovery the twisted space-times for $\;{\mathcal U}_{\kappa}({{\widehat{NH}}_{\pm}})$,
$\;{\mathcal U}_{\kappa}({{\widehat{G}}})$, $\;{\mathcal U}_{\kappa}({{{NH}}_{\pm}})$ and $\;{\mathcal U}_{\kappa}({{{G}}})$ quantum groups provided in articles  \cite{Daszkiewicz:2010bp}, \cite{twistnh} and \cite{dasz1} respectively\footnote{For general classification of all deformations of relativistic and nonrelativistic Hopf algebras see \cite{zak}, \cite{bri}.}. Besides, we find completely new quantum spaces associated with the Abelian twist factors containing additional generators $R_i$. In the case of finite value of parameter $\tau$ they take the form
\begin{equation}
[\;t,{ x}_{i}\;] = 0\;\;\;,\;\;\; [\;{ x}_{i},{ x}_{j}\;] = 
if_{\pm}\left(\frac{t}{\tau}\right)\theta_{ij}(x)
\;, \label{nhspace}
\end{equation}
with time-dependent  functions
$$f_+\left(\frac{t}{\tau}\right) =
f\left(\sinh\left(\frac{t}{\tau}\right),\cosh\left(\frac{t}{\tau}\right)\right)\;\;\;,\;\;\;
f_-\left(\frac{t}{\tau}\right) =
f\left(\sin\left(\frac{t}{\tau}\right),\cos\left(\frac{t}{\tau}\right)\right)\;,$$
$$\theta_{ij}(x) \sim \theta_{ij} = {\rm const}\;\; {\rm or}\;\;
\theta_{ij}(x) \sim \theta_{ij}^{k}x_k\;,$$
while  for cosmological constans $\tau$ approaching infinity, they look as follows
\begin{equation}
[\;{ x}_{\mu},{ x}_{\nu}\;] = i\alpha_{\mu\nu}^{\rho_1...\rho_n}{
x}_{\rho_1}...{ x}_{\rho_n}\;, \label{noncomm6}
\end{equation}
with $n=3,4,5$ and $6$. 

It should be noted, that the motivation for present studies are (at least) twofold. First of all, in accordance with the contraction procedures i)-v), we consider
 the Abelian twists of the largest known explicitly  nonrelativistic Hopf structure. Consequently, the obtained results permit to analyze the
classical as well as quantum particle models defined on the corresponding ("largest") twisted space-times (the similar investigations for simpler quantum groups have been performed in articles
\cite{Romero:2002ns}-\cite{Miao:2009jw}).

The paper is organized as follows. In second section fifteen   Abelian
classical $r$-matrices for twisted doubly enlarged
Newton-Hooke Hopf algebras are considered.  The corresponding fifteen
quantum space-times are provided in section 3, while  their $\tau
\to \infty$ contractions to the doubly enlarged Galilei spaces
are discussed in section 4. The final remarks are presented in the last
section.

\section{{{Twisted doubly enlarged  Newton-Hooke Hopf algebras}}}

In accordance with Drinfeld  twist procedure
\cite{twist1}-\cite{twist3}, the algebraic sector of twisted
doubly enlarged Newton-Hooke Hopf algebra $\;{\mathcal U}_{0}(\widehat{{\widehat{NH}}}_{\pm})$ remains
undeformed  (see (\ref{nnnga})), while
the   coproducts and antipodes  transform as follows (see formula (\ref{cop}))
\begin{equation}
\Delta _{0}(a) \to \Delta _{\cdot }(a) = \mathcal{F}_{\cdot }\circ
\,\Delta _{0}(a)\,\circ \mathcal{F}_{\cdot }^{-1}\;\;\;,\;\;\;
S_{\cdot}(a) =u_{\cdot }\,S_{0}(a)\,u^{-1}_{\cdot }\;,\label{fs}
\end{equation}
with $u_{\cdot }=\sum f_{(1)}S_0(f_{(2)})$ (we use Sweedler's notation
$\mathcal{F}_{\cdot }=\sum f_{(1)}\otimes f_{(2)}$).
Besides, it should be noted, that the twist factor
$\mathcal{F}_{\cdot } \in {\mathcal U}_{\cdot}(\widehat{\widehat{NH}}_{\pm}) \otimes
{\mathcal U}_{\cdot}(\widehat{\widehat{NH}}_{\pm})$
satisfies  the classical cocycle condition
\begin{equation}
{\mathcal F}_{{\cdot }12} \cdot(\Delta_{0} \otimes 1) ~{\cal
F}_{\cdot } = {\mathcal F}_{{\cdot }23} \cdot(1\otimes \Delta_{0})
~{\mathcal F}_{{\cdot }}\;, \label{cocyclef}
\end{equation}
and the normalization condition
\begin{equation}
(\epsilon \otimes 1)~{\cal F}_{{\cdot }} = (1 \otimes
\epsilon)~{\cal F}_{{\cdot }} = 1\;, \label{normalizationhh}
\end{equation}
with ${\cal F}_{{\cdot }12} = {\cal F}_{{\cdot }}\otimes 1$ and
${\cal F}_{{\cdot }23} = 1 \otimes {\cal F}_{{\cdot }}$.

It is well known, that the twisted algebra $\;{\mathcal U}_{\cdot}(\widehat{\widehat{NH}}_{\pm})$ can be described in terms of
so-called classical $r$-matrix $r\in {\mathcal U}_{\cdot}(\widehat{\widehat{NH}}_{\pm}) \otimes {\mathcal U}_{\cdot}(\widehat{\widehat{NH}}_{\pm})$,
which satisfies the  classical Yang-Baxter equation (CYBE)
\begin{equation}
[[\;r_{\cdot},r_{\cdot}\;] ] = [\;r_{\cdot 12},r_{\cdot13} +
r_{\cdot 23}\;] + [\;r_{\cdot 13}, r_{\cdot 23}\;] = 0\;,
\label{cybe}
\end{equation}
where   symbol $[[\;\cdot,\cdot\;]]$ denotes the Schouten bracket
and for $r = \sum_{i}a_i\otimes b_i$
$$r_{ 12} = \sum_{i}a_i\otimes b_i\otimes 1\;\;,\;\;r_{ 13} = \sum_{i}a_i\otimes 1\otimes b_i\;\;,\;\;
r_{ 23} = \sum_{i}1\otimes a_i\otimes b_i\;.$$

In this article  we consider fifteen
 Abelian twist-deformations of  doubly enlarged Newton-Hooke Hopf
algebra, described by the following $r$-matrices\footnote{$a\wedge b
= a\otimes b - b\otimes a$.}
\begin{eqnarray}
1)\;\;\;\;r_{\beta_1} &=&  \frac{1}{2}{\beta_1^{kl}} R_k \wedge
R_l\;\;\;\;\;\;\;\, [\;\beta_1^{kl} = -\beta_1^{lk}\;]\;,
\label{rmacierze01}\\&~~&\cr
2)\;\;\;\;r_{\beta_2} &=&  \frac{1}{2}{\beta_2^{kl}} R_k \wedge
F_l\;\;\;\;\;\;\;\, [\;\beta_2^{kl} = -\beta_2^{lk}\;]\;,
\label{rmacierze02}\\&~~&\cr
3)\;\;\;\;r_{\beta_3} &=&  \frac{1}{2}{\beta_3^{kl}} R_k \wedge
P_l\;\;\;\;\;\;\;\, [\;\beta_3^{kl} = -\beta_3^{lk}\;]\;,
\label{rmacierze03}\\&~~&\cr
4)\;\;\;\;r_{\beta_4} &=&  \frac{1}{2}{\beta_4^{kl}} K_k \wedge
R_l\;\;\;\;\;\;\;\, [\;\beta_4^{kl} = -\beta_4^{lk}\;]\;,
\label{rmacierze04}\\&~~&\cr
5)\;\;\;\;r_{\beta_5} &=&  \beta_5 R_m
\wedge M_{kl}\;\;\; [\;m,k,l - {\rm fixed},\;\;m \neq
k,l\;]\;,\label{rmacierzen00}\\&~~&\cr
6)\;\;\;\;r_{\beta_6} &=&  \frac{1}{2}{\beta_6^{kl}} F_k \wedge
F_l\;\;\;\;\;\;\;\, [\;\beta_6^{kl} = -\beta_6^{lk}\;]\;,
\label{macierze01}\\&~~&\cr
 7)\;\;\;\;r_{\beta_7} &=&
\frac{1}{2}\beta_7^{kl} F_k \wedge P_l\;\;\;\;\;\;\; [\;\beta_7^{kl}
= -\beta_7^{lk}\;]\;,\label{macierze100}\\&~~&\cr
8)\;\;\;\;r_{\beta_8} &=& \frac{1}{2}{\beta_8^{kl}} K_k \wedge
F_l\;\;\;\;\;\;\, [\;\beta_8^{kl} = -\beta_8^{lk}\;]\;,
\label{macierze200}\\&~~&\cr
 9)\;\;\;\;r_{\beta_9} &=&  \beta_9 F_m
\wedge M_{kl}\;\;\; [\;m,k,l - {\rm fixed},\;\;m \neq
k,l\;]\;,\label{macierzen00}\\&~~&\cr
 10)\;\;\;\;r_{\beta_{10}} &=&
\frac{1}{2}{\beta_{10}^{kl}} P_k \wedge P_l\;\;\;\;\;\;\;\,
[\;\beta_{10}^{kl} = -\beta_{10}^{lk}\;]\;, \label{macierze0}\\
&~~&\cr 11)\;\;\;\;r_{\beta_{11}} &=& \frac{1}{2}\beta_{11}^{kl} K_k \wedge
P_l\;\;\;\;\;\;\; [\;\beta_{11}^{kl} = -\beta_{11}^{lk}\;]\;,\\ &~~&\cr
12)\;\;\;\;r_{\beta_{12}} &=& \frac{1}{2}{\beta_{12}^{kl}} K_k \wedge
K_l\;\;\;\;\;\;\, [\;\beta_{12}^{kl} = -\beta_{12}^{lk}\;]\;,
\label{macierze}\\
&~~&\cr 13)\;\;\;\; r_{\beta_{13}} &=&  \beta_{13} K_m \wedge M_{kl}\;\;\;
[\;m,k,l - {\rm fixed},\;\;m \neq k,l\;]\;,\\ &~~&\cr 14)\;\;\;\;
r_{\beta_{14}} &=& \beta_{14} P_m \wedge M_{kl}\;\;\; [\;m,k,l - {\rm
fixed},\;\;m \neq
k,l\;]\;,\label{macierze1}\\
&~~&\cr 15)\;\;\;\;r_{\beta_{15}} &=& \beta_{15} M_{ij} \wedge
H\;.\label{macierzenn}
\end{eqnarray}
Due to  Abelian character of the above carriers (all of them arise
from the mutually commuting elements of the algebra), the
corresponding twist factors can be get in a  standard way
\cite{twist1}-\cite{twist3}, i.e. they take the form
\begin{eqnarray}
{\cal F}_{{\beta_k }} = \exp
\left(ir_{\beta_k}\right)\;\;\;;\;\;\;k=1,2,...,15\;.
\label{factors}
\end{eqnarray}
Let us  note that   first five  matrices include (new)
generators $R_i$, while the next ten  factors are the same as in the case of
acceleration-enlarged Newton-Hooke Hopf algebra considered
in \cite{Daszkiewicz:2010bp}. Of course, for all
deformation parameters $\beta_i$ approaching zero the discussed
above Hopf structures $\;{\mathcal U}_{\beta_i}(\widehat{\widehat{NH}}_{\pm})$ become classical, i.e. they become undeformed.

\section{{{Quantum doubly enlarged  Newton-Hooke  space-times}}}

Let us now turn to the deformed space-times corresponding to the
twist-deformations $1)$-$15)$ discussed in pervious section. They
are defined as the quantum representation spaces (Hopf modules) for
quantum doubly enlarged Newton-Hooke algebras, with action of
the deformed symmetry generators satisfying suitably deformed
Leibnitz rules \cite{bloch}-\cite{chi}.

The action of generators $M_{ij}$, $K_i$, $P_i$, $F_i$, $H$ and $R_i$ on a
Hopf module of functions depending on space-time coordinates
$(t,x_i)$ is given by
\begin{equation}
H\rhd f(t,\overline{x})=i{\partial_t}f(t,\overline{x})\;\;\;,\;\;\;
P_{i}\rhd f(t,\overline{x})=iC_{\pm} \left(\frac{t}{\tau}\right)
{\partial_i}f(t,\overline{x})\;, \label{a1}
\end{equation}
\begin{equation}
M_{ij}\rhd f(t,\overline{x}) =i\left( x_{i }{\partial_j} -x_{j
}{\partial_i} \right) f(t,\overline{x})\;\;\;,\;\;\; K_i\rhd
f(t,\overline{x}) =i\tau \,S_{\pm} \left(\frac{t}{\tau}\right)
{\partial_i} \,f(t,\overline{x})\;,\label{dsfa}
\end{equation}
\begin{equation}
F_i\rhd f(t,\overline{x})=\pm 2i\tau^2\left(C_{\pm}
\left(\frac{t}{\tau}\right) -1\right)
{\partial_i}f(t,\overline{x})\;,  \label{dsf}
\end{equation}
and
\begin{equation}
R_i\rhd f(t,\overline{x})= \pm 6i\tau^3\left(S_{\pm} \left(\frac{t}{\tau}\right)  - \frac{t}{\tau}\right)\;.\label{dsfff}
\end{equation}
Moreover, the $\star$-multiplication of arbitrary two functions  is
defined as follows
\begin{equation}
f(t,\overline{x})\star_{\beta_i} g(t,\overline{x}):=
\omega\circ\left(
 \mathcal{F}_{\beta_i}^{-1}\rhd  f(t,\overline{x})\otimes g(t,\overline{x})\right)
 \;,
\label{star}
\end{equation}
where symbol  $\mathcal{F}_{\beta_i}$ denotes the  twist factor (see
(\ref{factors})) corresponding to the proper doubly enlarged
Newton-Hooke Hopf algebra and $\omega\circ\left( a\otimes b\right) =
a\cdot b$.

In such a way  we get fifteen  quantum   space-times
\begin{eqnarray}
&1)&[\,t,x_a\,]_{{\star}_{\beta_1}} =0\;\;\;,\;\;\;
[\,x_a,x_b\,]_{{\star}_{\beta_1}}
= 36i\beta_1^{kl}\tau^6
\left(S_{\pm} \left(\frac{t}{\tau}\right)  - \frac{t}{\tau}\right)^2
 (\delta_{ak}\delta_{bl} -
\delta_{al}\delta_{bk})\;,\label{rspacetime1}\\&~~&  \cr
&2)&[\,t,x_a\,]_{{\star}_{\beta_2}} =0\;,\nonumber ~\\
&&[\,x_a,x_b\,]_{{\star}_{\beta_2}} =
12 i\beta_2^{kl}\tau^5
\left(S_{\pm} \left(\frac{t}{\tau}\right)  - \frac{t}{\tau}\right)
\left(C_{\pm}\left(\frac{t}{\tau}\right) -1\right)(\delta_{ak}\delta_{bl} -
\delta_{al}\delta_{bk})\;,\label{rspacetime2}\\&~~&  \cr
&3)&[\,t,x_a\,]_{{\star}_{\beta_3}} =0\;,\nonumber\\&~~&  \cr
&&[\,x_a,x_b\,]_{{\star}_{\beta_3}} =\pm 6 i\beta_3^{kl}\tau^3
\left(S_{\pm} \left(\frac{t}{\tau}\right)  - \frac{t}{\tau}\right)C_{\pm}
\left(\frac{t}{\tau}\right)(\delta_{ak}\delta_{bl} -
\delta_{al}\delta_{bk})\;,\label{rspacetime3}\\&~~&  \cr
&4)&[\,t,x_a\,]_{{\star}_{\beta_4}} =0\;,\nonumber ~\\&~~&  \cr
&&[\,x_a,x_b\,]_{{\star}_{\beta_4}} =
\pm 6i\beta_4^{kl}\tau^4 \left(S_{\pm} \left(\frac{t}{\tau}\right)  - \frac{t}{\tau}\right)S_{\pm}
\left(\frac{t}{\tau}\right)(\delta_{ak}\delta_{bl} -
\delta_{al}\delta_{bk})\;,
\label{rspacetime4}\\&~~&
\cr &5)&[\,t,x_a\,]_{{\star}_{\beta_5}} =0\;,\\&~~&  \cr
&&[\,x_a,x_b\,]_{{\star}_{\beta_5}} =\pm 12i\beta_5\tau^3 \left(S_{\pm} \left(\frac{t}{\tau}\right)  - \frac{t}{\tau}\right)
\left[\;\delta_{ma}(x_k\delta_{bl} - x_{l}\delta_{bk}) -
\delta_{mb}(x_k\delta_{al} -
x_{l}\delta_{ik})\;\right]\label{rspacetime5}\\&~~&  \cr
&6)&[\,t,x_a\,]_{{\star}_{\beta_6}} =0\;\;\;,\;\;\;
[\,x_a,x_b\,]_{{\star}_{\beta_6}}
=4i\beta_6^{kl}\tau^4\left(C_{\pm}\left(\frac{t}{\tau}\right)
-1\right)^2 (\delta_{ak}\delta_{bl} -
\delta_{al}\delta_{bk})\;,\label{spacetime1}\\&~~&  \cr
&7)&[\,t,x_a\,]_{{\star}_{\beta_7}} =0\;,\nonumber ~\\
&&[\,x_a,x_b\,]_{{\star}_{\beta_7}} =\pm i\beta_7^{kl}\tau^2
\left(C_{\pm}\left(\frac{t}{\tau}\right) -1\right)C_{\pm}
\left(\frac{t}{\tau}\right)(\delta_{ak}\delta_{bl} -
\delta_{al}\delta_{bk})\;,\label{spacetime2}\\&~~&  \cr
&8)&[\,t,x_a\,]_{{\star}_{\beta_8}} =0\;,\nonumber\\&~~&  \cr
&&[\,x_a,x_b\,]_{{\star}_{\beta_8}} =\pm i\beta_8^{kl}\tau^3
\left(C_{\pm}\left(\frac{t}{\tau}\right) -1\right)S_{\pm}
\left(\frac{t}{\tau}\right)(\delta_{ak}\delta_{bl} -
\delta_{al}\delta_{bk})\;,\label{spacetime3}\\&~~&  \cr
&9)&[\,t,x_a\,]_{{\star}_{\beta_9}} =0\;,\nonumber ~\\&~~&  \cr
&&[\,x_a,x_b\,]_{{\star}_{\beta_9}} =\pm 4i\beta_9\tau^2
\left(C_{\pm}\left(\frac{t}{\tau}\right)
-1\right)\left[\;\delta_{ma}(x_k\delta_{bl} - x_{l}\delta_{bk}) -
\delta_{mb}(x_k\delta_{al} -
x_{l}\delta_{ik})\;\right]\;\;\;\;\;\;\;\;\label{spacetime4}\\&~~&
\cr &10)&[\,t,x_a\,]_{{\star}_{\beta_{10}}} =0\;\;\;,\;\;\;
[\,x_a,x_b\,]_{{\star}_{\beta_{10}}} =i\beta_{10}^{kl}\,C_{\pm}^2
\left(\frac{t}{\tau}\right)(\delta_{ak}\delta_{bl} -
\delta_{al}\delta_{bk})\;,\label{spacetime5}\\&~~&  \cr
&11)&[\,t,x_a\,]_{{\star}_{\beta_{11}}} =0\;\;\;,\;\;\;
[\,x_a,x_b\,]_{{\star}_{\beta_{11}}} =i\beta_{11}^{kl}\tau\, C_{\pm}
\left(\frac{t}{\tau}\right)S_{\pm}
\left(\frac{t}{\tau}\right)(\delta_{ak}\delta_{bl} -
\delta_{al}\delta_{bk})\;,\label{spacetime6}\\&~~&  \cr
&12)&[\,t,x_a\,]_{{\star}_{\beta_{12}}} =0\;\;\;,\;\;\;
[\,x_a,x_b\,]_{{\star}_{\beta_{12}}} =i\beta_{12}^{kl}\tau^2\, S_{\pm}^2
\left(\frac{t}{\tau}\right)(\delta_{ak}\delta_{bl} -
\delta_{al}\delta_{bk})\;,\label{spacetime7}\\&~~&  \cr
&13)&[\,t,x_a\,]_{{\star}_{\beta_{13}}} =0\;,\nonumber ~\\&~~&  \cr
&&[\,x_a,x_b\,]_{{\star}_{\beta_{13}}} =2i\beta_{13}\tau\, S_{\pm}
\left(\frac{t}{\tau}\right)\left[\;\delta_{ma}(x_k\delta_{bl} -
x_{l}\delta_{bk}) - \delta_{mb}(x_k\delta_{al} -
x_{l}\delta_{ak})\;\right]\;,\label{spacetime8}\\&~~&  \cr
&14)&[\,t,x_a\,]_{{\star}_{\beta_{14}}} =0\;,\nonumber ~\\&~~&  \cr
&&[\,x_a,x_b\,]_{{\star}_{\beta_{14}}} =2i\beta_{14}\,C_{\pm}
\left(\frac{t}{\tau}\right)\left[\;\delta_{ma}(x_k\delta_{bl} -
x_{l}\delta_{bk}) - \delta_{mb}(x_k\delta_{al} -
x_{l}\delta_{ak})\;\right]\;,\label{spacetime9a}
\\&~~&  \cr
&15)&[\,t,x_a\,]_{{\star}_{\beta_{15}}} =2i\beta_{15}
\left[\;\delta_{ia}x_j - x_{i}\delta_{ja}\;\right] \;\;\;,\;\;\;
[\,x_a,x_b\,]_{{\star}_{\beta_{15}}} =0 \;,\label{spacetime9}
\end{eqnarray}
associated with  matrices $1)$-$15)$, respectively.

 Let us note that due to the form of functions
$C_{\pm} [\frac{t}{\tau}]$ and $S_{\pm} [\frac{t}{\tau}]$ the
spatial noncommutativities 1)-14) are expanding or periodic in time
respectively. Moreover, all of them introduce classical time and
quantum spatial directions. The last type of  space-time
noncommutativity provides the quantum time and classical spatial
variables. It should be also noted that in the contraction limit i) and v)
we get the twisted acceleration-enlarged Newton-Hooke and usual Newton-Hooke
spaces, respectively.

Of course,  for all  deformation parameters $\beta_i$ approaching
zero, the above quantum  space-times  become commutative.

\section{{{Twisted doubly enlarged  Galilei Hopf algebras - the $\bf \tau \to \infty$ limit (contraction v))}}}

In this section we provide  twisted doubly enlarged Galilei
Hopf algebras $\,{\mathcal U}_{\beta_i}(\widehat{\widehat{{G}}})$ and
 corresponding quantum space-times, as the $\tau \to \infty$
limit of Hopf structures discussed in pervious sections. In such a
limit the commutation relations (\ref{nnnga}) become
$\tau$-independent, i.e. we neglect the impact of the cosmological
time scale $\tau$ on the structure of the considered Hopf algebras.

First of all, we perform the contraction limit $\tau \to \infty$ of
the formulas (\ref{nnnga}) and
(\ref{rmacierze01})-(\ref{macierzenn}). Consequently, the
corresponding classical $r$-matrices remain the same as
(\ref{rmacierze01})-(\ref{macierzenn}), while the algebraic sector of
all considered $\,{\mathcal U}_{\beta_i}(\widehat{\widehat{{G}}})$
algebras takes the form
\begin{eqnarray}
&&\left[\, M_{ij},M_{kl}\,\right] =i\left( \delta
_{il}\,M_{jk}-\delta _{jl}\,M_{ik}+\delta _{jk}M_{il}-\delta
_{ik}M_{jl}\right)\;\; \;, \;\;\; \left[\, H,P_i\,\right] =0
 \;,  \notag \\
&~~&  \cr &&\left[\, M_{ij},K_{k}\,\right] =i\left( \delta
_{jk}\,K_i-\delta _{ik}\,K_j\right)\;\; \;, \;\;\;\left[
\,M_{ij},P_{k }\,\right] =i\left( \delta _{j k }\,P_{i }-\delta _{ik
}\,P_{j }\right) \;, \label{gggali}
\\
&~~&  \cr &&\left[ \,M_{ij},H\,\right] =\left[ \,K_i,K_j\,\right] =
\left[ \,K_i,P_{j }\,\right] =0\;\;\;,\;\;\;\left[ \,K_i,H\,\right]
=-iP_i\;\;\;,\;\;\;\left[ \,P_{i },P_{j }\,\right] = 0\;,\nonumber\\
&~~&  \cr &&\left[\, F_i,F_j\,\right] =\left[\, F_i,P_j\,\right]
=\left[\, F_i,K_j\,\right] =0\;\; \;, \;\;\;\left[\,
M_{ij},F_{k}\,\right] =i\left( \delta _{jk}\,F_i-\delta
_{ik}\,F_j\right) \;,\nonumber\\
&~~&  \cr
&&\left[\, R_i,R_j\,\right] =\left[\, R_i,P_j\,\right]
=\left[\, R_i,K_j\,\right] =\left[\, R_i,F_j\,\right] =0\;\; \;, \;\;\;\left[\,
M_{ij},R_{k}\,\right] =i\left( \delta _{jk}\,R_i-\delta
_{ik}\,R_j\right) \;,\nonumber\\
&~~&  \cr &&~~~~~~~~~~~~~~~~~~~~~~~\left[\,
H,F_{i}\,\right] =2iK_i  \;\;\;,\;\;\;      \left[\,
H,R_{i}\,\right] =3iF_i\;.\nonumber
\end{eqnarray}
The corresponding coproduct sectors  can be get by application of
the formulas (\ref{fs}) and (\ref{factors}).

Let us now turn to the corresponding quantum nonrelativistic
space-times. One can check (see $\tau \to \infty$ limit of the
formulas (\ref{rspacetime1})-(\ref{spacetime9})) that they look as
follows\footnote{It should be noted that the commutation relations
(\ref{rgali1})-(\ref{gali9}) can be also derived  with use of the
formula (\ref{star}) and differential representation of
doubly enlarged Galilei generators.}
\begin{eqnarray}
&1)&[\,t,x_a\,]_{{\star}_{\beta_1}} =0\;\;\;,\;\;\;
[\,x_a,x_b\,]_{{\star}_{\beta_1}}
=i\beta_1^{kl}\,t^6\,(\delta_{ak}\delta_{bl} - \delta_{al}\delta_{bk})\;,\label{rgali1}\\
&~~&\cr &2)&[\,t,x_a\,]_{{\star}_{\beta_2}} =0\;\;\;,\;\;\;
[\,x_a,x_b\,]_{{\star}_{\beta_2}}
=i\beta_2^{kl}\,t^5\,(\delta_{ak}\delta_{bl} - \delta_{al}\delta_{bk})\;,\label{rgali2}\\
&~~&\cr &3)&[\,t,x_a\,]_{{\star}_{\beta_3}} =0\;\;\;,\;\;\;
[\,x_a,x_b\,]_{{\star}_{\beta_3}} =i\beta_3^{kl}\,t^3\,
(\delta_{ak}\delta_{bl} -
\delta_{al}\delta_{bk})\;,\label{rgali3}\\
&~~&\cr &4)&[\,t,x_a\,]_{{\star}_{\beta_4}} =0\;\;\;,\;\;\;
  [\,x_a,x_b\,]_{{\star}_{\beta_4}} =i\beta_4\,t^4\,
(\delta_{ak}\delta_{bl} - \delta_{al}\delta_{bk})
\;,\label{rgali4}\\
&~~&\cr &5)&[\,t,x_a\,]_{{\star}_{\beta_5}} =0\;,\nonumber
~\\&~~&\cr &&
[\,x_a,x_b\,]_{{\star}_{\beta_5}}
=2i\beta_5^{kl}t^3\left[\;\delta_{ma}(x_k\delta_{bl} - x_{l}\delta_{bk}) -
\delta_{mb}(x_k\delta_{al} -
x_{l}\delta_{ak})\;\right]\;,\label{rgali5}\\
&~~&\cr
&6)&[\,t,x_a\,]_{{\star}_{\beta_6}} =0\;\;\;,\;\;\;
[\,x_a,x_b\,]_{{\star}_{\beta_6}}
=i\beta_6^{kl}\,t^4\,(\delta_{ak}\delta_{bl} - \delta_{al}\delta_{bk})\;,\label{gali1}\\
&~~&\cr &7)&[\,t,x_a\,]_{{\star}_{\beta_7}} =0\;\;\;,\;\;\;
[\,x_a,x_b\,]_{{\star}_{\beta_7}}
=\frac{i}{2}\beta_2^{kl}\,t^2\,(\delta_{ak}\delta_{bl} - \delta_{al}\delta_{bk})\;,\label{gali2}\\
&~~&\cr &8)&[\,t,x_a\,]_{{\star}_{\beta_8}} =0\;\;\;,\;\;\;
[\,x_a,x_b\,]_{{\star}_{\beta_8}} =\frac{i}{2}\beta_8^{kl}\,t^3\,
(\delta_{ak}\delta_{bl} -
\delta_{al}\delta_{bk})\;,\label{gali3}\\
&~~&\cr &9)&[\,t,x_a\,]_{{\star}_{\beta_9}} =0\;,\nonumber
~\\&~~&\cr && [\,x_a,x_b\,]_{{\star}_{\beta_9}} =2i\beta_4\,t^2\,
\left[\;\delta_{ma}(x_k\delta_{bl} - x_{l}\delta_{bk}) -
\delta_{mb}(x_k\delta_{al} -
x_{l}\delta_{ak})\;\right]\;,\label{gali4}\\
&~~&\cr &10)&[\,t,x_a\,]_{{\star}_{\beta_{10}}} =0\;\;\;,\;\;\;
[\,x_a,x_b\,]_{{\star}_{\beta_{10}}}
=i\beta_{10}^{kl}(\delta_{ak}\delta_{bl} - \delta_{al}\delta_{bk})\;,\label{gali5}\\
&~~&\cr &11)&[\,t,x_a\,]_{{\star}_{\beta_{11}}} =0\;\;\;,\;\;\;
[\,x_a,x_b\,]_{{\star}_{\beta_{11}}}
=i\beta_{11}^{kl}\,t\,(\delta_{ak}\delta_{bl} - \delta_{al}\delta_{bk})\;,\label{gali6}\\
&~~&\cr &12)&[\,t,x_a\,]_{{\star}_{\beta_{12}}} =0\;\;\;,\;\;\;
[\,x_a,x_b\,]_{{\star}_{\beta_{12}}} =i\beta_{12}^{kl}\,t^2\,
(\delta_{ak}\delta_{bl} - \delta_{al}\delta_{bk})\;,\label{gali7}\\
&~~&\cr &13)&[\,t,x_a\,]_{{\star}_{\beta_{13}}} =0\;,\nonumber\\&~~&\cr
&&[\,x_a,x_b\,]_{{\star}_{\beta_{13}}} =2i\beta_{13}\,t\,
\left[\;\delta_{ma}(x_k\delta_{bl} - x_{l}\delta_{bk}) -
\delta_{mb}(x_k\delta_{al} - x_{l}\delta_{ak})\;\right]\;,\label{gali8}\\
&~~&\cr &14)&[\,t,x_a\,]_{{\star}_{\beta_{14}}} =0\;,\nonumber\\&~~&\cr
&&[\,x_a,x_b\,]_{{\star}_{\beta_{14}}}
=2i\beta_{14}\,\left[\;\delta_{ma}(x_k\delta_{bl} - x_{l}\delta_{bk}) -
\delta_{mb}(x_k\delta_{al} -
x_{l}\delta_{ak})\;\right]\;,\label{gali9a}\\&~~&\cr
&15)&[\,t,x_a\,]_{{\star}_{\beta_{15}}} =2i\beta_{15}
\left[\;\delta_{ia}x_j - x_{i}\delta_{ja}\;\right] \;\;\;,\;\;\;
[\,x_a,x_b\,]_{{\star}_{\beta_{15}}} =0\;,\label{gali9}
\end{eqnarray}
in the case of $\,{\mathcal U}_{\beta_1}(\widehat{\widehat{{{ G}}}}),
\dots, {\mathcal U}_{\beta_{15}}(\widehat{\widehat{{{ G}}}})$ Hopf
algebras respectively.

Obviously, for all deformation parameters $\beta_i$ approaching zero
the above Hopf algebras become classical, while the corresponding
quantum space-times - commutative.

\section{Final remarks}

In this article we consider fifteen  Abelian twist-deformations of
doubly enlarged Newton-Hooke Hopf algebras. Besides, we
demonstrate that as in the case of  twist-deformed
acceleration-enlarged Newton-Hooke Hopf algebra, the corresponding spaces can be  periodic
and expanding in time for $\,{\mathcal U}_{\beta_i}(\widehat{\widehat{
NH}}_{-})$ and $\,{\mathcal U}_{\beta_i}(\widehat{\widehat{ NH}}_{+})$ quantum
groups respectively. In $\tau \to \infty$  limit we also discover
new twisted doubly enlarged Galilei  space-times (\ref{rgali1})-(\ref{gali9}).

It should be noted that  present studies can be extended in various
ways.  First of all, one can find the dual Hopf structures
$\,{\mathcal D}_{\beta_i}(\widehat{\widehat{NH}}_{\pm})$ with the use of FRT
procedure \cite{frt} or by canonical quantization of the
corresponding Poisson-Lie structures \cite{poisson}. Besides, as it
was already mentioned in Introduction, one should ask about the
basic dynamical models corresponding to the doubly enlarged
Newton-Hooke space-times
(\ref{rspacetime1})-(\ref{spacetime9}).   Finally, one can also consider more
complicated (non-Abelian) twist deformations of
doubly enlarged Newton-Hooke Hopf algebras, i.e. one can find
the twisted coproducts, corresponding noncommutative space-times and
dual Hopf structures. Such  problems are now under consideration.

\section*{Acknowledgments}
The author would like to thank J. Lukierski
for valuable discussions. This paper has been financially  supported  by Polish
NCN grant No 2011/01/B/ST2/03354.

\end{document}